\documentclass{article}

\usepackage[preprint]{neurips_2026}

\usepackage[utf8]{inputenc}
\usepackage[T1]{fontenc}
\usepackage{hyperref}
\usepackage{url}
\usepackage{booktabs}
\usepackage{amsfonts}
\usepackage{nicefrac}
\usepackage{microtype}
\usepackage{xcolor}
\usepackage{graphicx}
\usepackage{amsmath}
\usepackage{amssymb}
\usepackage{multirow}
\usepackage{array}
\usepackage{subcaption}
\usepackage{wrapfig}
\usepackage{float}

\title{Reviewer Scores Are Not Comparable Across Research Areas in ML Peer Review}

\author{
  Binyan Xu$^{1}$,
  Xilin Dai$^{2}$,
  Fan Yang$^{1}$,
  and Kehuan Zhang$^{1}$ \\[4pt]
  $^{1}$The Chinese University of Hong Kong, Hong Kong, China \\
  $^{2}$Zhejiang University, Hangzhou, China \\
  \texttt{\{binyxu,yf020,khzhang\}@ie.cuhk.edu.hk, xilin2023@zju.edu.cn}
}

\begin{document}

\maketitle

\begin{abstract}
Peer review at ML conferences increasingly uses reviewer scores as the primary decision instrument. Yet scores are produced by topic-specific reviewer pools. Different topics are evaluated by different, only partly overlapping sets of reviewers, and the process lacks a designed procedure that empirically equates their scales. Using ICLR 2021--2026 data covering 50,289 papers and a cleaned taxonomy of 219 research topics, we find that topic improves acceptance prediction after controlling for score and year (global likelihood-ratio $p=5.1\times10^{-3}$; paper-clustered $p=1.8\times10^{-2}$). At a fixed recorded score, the largest ratio among our reported topic-band summaries is 8$\times$; it is an observed maximum, not a typical effect. In 2026 at score $\approx5.0$, the median absolute pairwise topic gap is 6.18 percentage points and the 90th-to-10th-percentile spread is 15.29 points. These differences establish that one nominal reviewer score has no stable decision meaning across research topics; they do not imply equal latent quality at equal scores or that every topic effect is unjustified. To our knowledge, this is the first large-scale audit to establish this failure through direct same-score comparisons and a joint conditional test. We test observable predictions of scoring culture, reviewer expertise, simple recorded-subscore reweighting, and quality dilution. None restores a common interpretation of the recorded score. We therefore diagnose a system-level measurement-design problem and call for topic-stratified, score-conditional reporting, explicit area criteria, and complementary calibrated signals.
\end{abstract}

%% ============================================================
\section{Introduction}
\label{sec:intro}
%% ============================================================

Machine learning conference venues have scaled dramatically: ICLR processed 2,978 effective submissions in 2021 and 18,790 in 2026, a 6.3$\times$ increase in 5 years~\citep{papercopilot2024}. At this scale, the review process depends on a shared scoring instrument: thousands of reviewers assign numerical ratings, and area chairs interpret those ratings when making accept/reject decisions. If one nominal scale is to serve as a venue-wide decision signal, its values need either a stable decision interpretation or an explicit calibration across reviewer pools. We test this decision-scale invariance by asking whether the same recorded score maps to the same acceptance probability across topics. This test does \emph{not} assume that equal scores represent equal latent paper quality. Whether the score has a common decision meaning has received little systematic scrutiny~\citep{shah2022challenges,tran2020open}.

The observed failure of decision-scale invariance has concrete procedural consequences: among papers with the same recorded score, topic supplies information about the decision that the score does not encode. Our data cannot determine which of two papers has higher latent quality, nor do they imply that every area-specific adjustment is unfair. The narrower conclusion is that the recorded score alone cannot distinguish justified calibration or additional AC information from unexplained topic drift. This matters because the conference presents one numerical scale while leaving any topic-specific decoder implicit and therefore difficult to audit~\citep{lipton2019troubling}.

\textbf{We argue that ML conference peer review suffers from a measurement design failure.}
The primary instrument, mean reviewer score, has no stable observed decision mapping across research areas. By \emph{topic-specific reviewer pools}, we mean that papers in different topics are evaluated by different, only partly overlapping groups of reviewers; we do not estimate the causal effect of assigning any particular reviewer. Such assignment does not by itself prove incomparability, but it makes cross-pool equivalence a property that requires explicit design and verification. If local score meanings differ, calibration is needed; if ACs use valid information beyond the score, the recorded score is incomplete. In either case, the public decision instrument lacks an explicit, auditable cross-topic interpretation. Our diagnosis concerns this system-level design and makes no claim of individual bias or carelessness.

We present three levels of evidence across 50,289 ICLR papers from 2021 to 2026. The first is descriptive: acceptance-rate disparities across topics. The second is conditional: a joint topic block improves acceptance prediction beyond recorded score and year. The third reads acceptance rates directly within narrow score bands: the largest reported ratio is 8$\times$, while representative dispersion summaries are substantially smaller. We then test specific observable predictions of four possible explanations: scoring culture, reviewer expertise, simple recorded-subscore reweighting, and quality dilution. These diagnostics narrow simple versions of the explanations but do not rule out every latent-quality or area-specific decision model. We conclude with remedies at two levels: transparency and explicit criteria for area-level drift, plus individual-level calibration signals as supplements.

As an important qualifier, our analysis is observational and confined to ICLR because it provides unusually complete public scores and decisions. We do not identify the causal effect of topic assignment or adjudicate the true quality of individual papers. Whether the same pattern occurs at NeurIPS, ICML, or other venues is unknown. Appendix~\ref{app:neurips} documents an attempted NeurIPS 2024 audit whose opt-in public sample is too selected for an external-validity conclusion. Complete internal data would allow those venues to run the same audit; the incomplete public data do not establish either presence or absence of the effect.

%% ============================================================
\section{Evidence of Score-Acceptance Decoupling}
\label{sec:evidence}
%% ============================================================

Our analysis draws on ICLR 2021--2026 paper records from the PaperCopilot repository~\citep{papercopilot2024}, covering 50,289 papers with at least one reviewer score. BERTopic~\citep{grootendorst2022bertopic} yields a cleaned exclusive taxonomy of 219 topics. For fine-grained, overlapping retrieval, deterministic regular expressions derived from the candidate topic labels allow a paper to belong to multiple topics; the main forest plot retains 146 topics meeting its sample-size filters. Appendix~\ref{app:topic_stats} reconciles the topic counts and documents retrieval. This taxonomy spans all six years and is finer than the 21 official \texttt{primary\_area} labels (available only from 2024). An \emph{effective submission} is a paper accepted, rejected, or withdrawn with at least 3 reviewers' confidence scores, excluding desk-reject-equivalent withdrawals. Computed acceptance rates match official ICLR figures within 1 percentage point across all years. Each paper carries per-reviewer \texttt{rating} (1--10), \texttt{confidence} (1--5), \texttt{soundness}, \texttt{contribution}, and \texttt{presentation} scores; we use mean reviewer rating as the recorded decision signal, not as a cross-area measure of latent quality. Per-topic 50\% logistic thresholds are descriptive summaries of score-to-decision curves; the narrow-band comparisons in Section~\ref{sec:inversion} require no logistic functional form. Prior work documents stochasticity~\citep{langford2015arbitrariness,cortes2021inconsistency,pier2018low}, inter-reviewer inconsistency~\citep{shah2022challenges,stelmakh2019biases}, and scale-driven arbitrariness~\citep{beygelzimer2023neurips} in peer review.

\subsection{Descriptive: Acceptance Rate Disparities Across Research Topics}
\label{sec:descriptive}

Figure~\ref{fig:systematic} (Appendix~\ref{app:systematic}) shows a forest plot of the acceptance rate for each of 146 qualifying topics (2021--2026, $\geq$250 matched papers, borderline $n\geq30$) with 95\% Wilson confidence intervals~\citep{wilson1927probable}. The spread from lowest (recognition: 18.3\%) to highest (radiance fields: 46.0\%) is 27.8 percentage points, with an overall pooled acceptance rate of 30.4\%.

Testing each of the 146 topic rates against the pooled rate as one family gives 59 nominal $p<0.05$ results. Of these, 49 remain significant under Benjamini--Hochberg FDR control, 31 under dependence-robust Benjamini--Yekutieli FDR control, and 19 under Holm family-wise correction (all at 0.05; Appendix~\ref{app:multiplicity}). Examples surviving all three corrections include lower-AR medical-imaging and graph-contrastive topics and the higher-AR robot-manipulation topic. Individual-topic claims use the corrected results.

\textbf{A caveat on 2024-only primary areas.} Three labels used only in 2024 (RepL/CV, General ML, Societal/Fairness) show the most extreme disadvantages; results for these areas may partly reflect definition ambiguity. The BERTopic topic-level analyzes are content-derived and do not use these author-assigned labels.

\textbf{Controlling for reviewer scores via logistic regression.}
The descriptive AR differences could in principle reflect genuine quality differences between submissions to different topics. Our baseline is $\operatorname{logit}P(A)=\alpha+\beta S+\gamma Y$; the nested model adds topic indicators. Using 113,290 paper--topic observations from 44,177 unique papers, the added topic block gives $\chi^2=341.2$ ($df=277$, $p=5.1\times10^{-3}$), with paper-clustered inference giving $p=1.8\times10^{-2}$. A stricter replication keeps one row per paper, enters all 146 qualifying overlapping topic indicators jointly, and allows a separate cubic score curve in each year; the topic block remains significant ($\chi^2=180.4$, $df=146$, $p=2.79\times10^{-2}$). An expanded audit finds that observed disagreement, confidence, review intensity, interaction proxies, and available quality subscores do not attenuate this joint topic result. Appendix~\ref{app:observable-adjustment} gives the complete analysis. \textbf{Thus topic improves acceptance prediction beyond the recorded score, year, and the observed auxiliary signals.} This conditional result concerns the decision mapping; it does not establish equal latent quality at equal scores.

\subsection{The Acceptance Bar Itself Is Different}
\label{sec:threshold}

A skeptic can offer two distinct explanations: \emph{trending topics may attract stronger submissions, and different communities may use the score scale differently}. The second explanation implies that equal numerical scores need not represent equal latent quality; generosity by itself does not make a higher score more meaningful.

To summarize the observed score-to-decision mapping, we fit a logistic curve $P(\text{accept}) = \sigma(\beta_0 + \beta_1 \cdot \text{score})$ on the aligned score scale for each topic and compute $\hat{\tau} = -\hat{\beta}_0/\hat{\beta}_1$, the aligned score at which predicted acceptance reaches 50\%. Figure~\ref{fig:threshold} (Appendix~\ref{app:threshold}) shows this descriptive threshold by topic:

\begin{itemize}
\vspace{-0.1cm}
\item \textbf{Lowest threshold (chain-of-thought/thinking topics)}: $\hat{\tau} = 4.938$ (easiest to get in)
\vspace{-0.05cm}
\item \textbf{Highest threshold (activation functions/relu)}: $\hat{\tau} = 5.751$ (hardest to get in)
\vspace{-0.05cm}
\item \textbf{Spread: 0.812 points on the aligned reviewer scale}
\vspace{-0.1cm}
\end{itemize}

The fitted threshold summarizes the observed decision curve; it is not a model of latent quality, and the one-dimensional logistic form is not needed for the narrow-band result below. Simple score inflation with an otherwise common decision mapping predicts an upward shift of the threshold, whereas several topics described as trending have lower fitted thresholds. More complex combinations can also generate this pattern. One example combines stronger submissions with stricter reviewers. Threshold variation alone does not identify the explanation.

\textbf{Descriptive score--acceptance decoupling.}
Areas with higher acceptance rates often have lower mean recorded scores among accepted papers. This negative correlation shows that observed scores alone do not summarize acceptance, but it does not rule out cross-area quality differences: better submissions combined with stricter scale use are a valid counterexample. We therefore treat this correlation as descriptive support only.

\textbf{A simple reviewer-expertise prediction is unsupported.}
If high-threshold topics are stricter solely because their reviewers are more expert, and self-reported confidence is an adequate expertise proxy, threshold and mean confidence should be positively correlated. We find $r=-0.102$ ($p=0.626$). The rating-to-confidence ratio correlates positively with acceptance rate ($r=+0.611$, $p=0.0012$). Confidence is an imperfect proxy, so these results do not establish equivalent expertise across topics. They also do not support this simple recorded-confidence explanation.

\subsection{Same Recorded Score, Different Acceptance Mapping Across Topics}
\label{sec:inversion}

The threshold analysis uses a logistic summary. Here we present direct narrow-band comparisons that do not assume a parametric decision curve (Appendix~\ref{app:samescorear}).

\textbf{Finding 1: At identical recorded reviewer scores, the observed maximum ratio reaches 8$\times$.}
For each of five score points ($s \in \{4.0, 4.5, 5.0, 5.5, 6.0\}$), we identify all topics with at least 50 papers in the band $[s-0.125,\, s+0.125]$, compute each topic's acceptance rate in that band, and compare the top-5 and bottom-5 topics.
Figure~\ref{fig:samescorear} shows the results across all five score points.
The gap is consistent and large at every level:

\begin{table}[t]
\setlength{\aboverulesep}{0.2ex}\setlength{\belowrulesep}{0.2ex}
\centering
\begin{small}
\caption{Acceptance rate spread across topics at each score band (all 6 years, $\geq$50 papers per topic in band). ``Top-5 / Bot-5 avg $n$'' = mean papers per topic in the two groups.}
\label{tab:samescore}
\smallskip
\begin{tabular}{cccccccc}
\toprule
Score & Topics & Overall AR & Top-5 AR & Bot-5 AR & Gap & Ratio & Top/Bot avg $n$ \\
\midrule
4.0 & 76 & 5.3\% & 11.6\% & 0.0\% & 12 pp & N/A & 63 / 75 \\
4.5 & 62 & 15.4\% & 29.0\% & 3.6\% & 25 pp & 8.0$\times$ & 82 / 99 \\
5.0 & 55 & 29.6\% & 52.6\% & 6.6\% & 46 pp & 8.0$\times$ & 84 / 72 \\
5.5 & 40 & 48.0\% & 69.6\% & 23.8\% & 46 pp & 2.9$\times$ & 93 / 75 \\
6.0 & 36 & 73.5\% & 87.0\% & 49.1\% & 38 pp & 1.8$\times$ & 67 / 59 \\
\bottomrule
\end{tabular}
\end{small}
\vspace{-0.6cm}
\end{table}

At score $\approx5.0$, the top-five topic average is 52.6\% and the bottom-five average is 6.6\%; at score $\approx5.5$, the corresponding averages are 69.6\% and 23.8\%. The 8$\times$ ratio is the largest observed ratio among the reported summaries, not a typical topic difference. As a representative within-year summary that avoids pooling conference years, in 2026 at score $\approx5.0$ the median absolute pairwise gap across 47 qualifying topics is 6.18 percentage points and the 90th-to-10th-percentile spread is 15.29 points. The direct rate spread is present at every reported score level and is largest near the middle of the scale, where decisions are least deterministic from scores alone.

\textbf{Supplementary analysis: a compounding mechanism.}
The same-score-band result above requires no grouping of topics. As a complementary illustration of the mechanism, Figure~\ref{fig:doubledisadvantage} splits topics into high-AR and low-AR groups and shows that the disparity operates at two stages.\footnote{Top-20 and bottom-20 groups are defined by overall six-year AR ($n\geq50$). The grouping is correlated with AR by construction; this figure illustrates the mechanism. The primary evidence is Table~\ref{tab:samescore}, which requires no grouping.}

\emph{Layer 1: Lower reviewer scores.}
The bottom-20 AR topics receive mean normalized reviewer scores of 4.49, compared to 5.11 for the top-20 AR topics ($\Delta = {+}0.622$, $t=9.56$, $p<0.001$).
Topics that are ultimately accepted less often are also scored lower by reviewers from the outset.

\emph{Layer 2: Lower acceptance at every score level.}
Even conditioning on reviewer score, the low-AR group is accepted less at every score band: +7\,pp gap at score $\approx$4.0, +9\,pp at 4.5, +13\,pp at 5.0, +21\,pp at 5.5, +12\,pp at 6.0.
The two layers compound in this grouped illustration. The grouping itself is circular, so this pattern does not establish independent effects on reviewer evaluation and AC decisions.

Rapid growth could lead communities to change their standards as frontiers such as multi-agent skill distillation expand~\citep{xu2026multi}. This account is compatible with topic-specific score meanings. It therefore reinforces the need to record and audit any calibration used to interpret the common numerical score.

\textbf{What area-specific thresholds can and cannot explain.}
Area-specific thresholds can produce different decisions at the same score and therefore can explain within-score-band gaps. If such thresholds are a deliberate response to different local scales or valid extra information, they constitute a topic-specific decoder for the common numerical score. The measurement-design concern is that this decoder is neither recorded nor auditable, so the public score alone cannot reveal whether the adjustment is justified. A narrower \emph{rate-control} hypothesis assumes that thresholds are chosen mainly to make area acceptance rates similar. It does not account for the observed 18.3--46.0\% topic-rate range. Thus we do not rule out rational area-specific thresholds; we argue that their necessity demonstrates the incompleteness of the nominally common score.

\textbf{A simple quality-dilution prediction is unsupported.}
A third alternative holds that rapidly growing areas mechanically receive more weak submissions. Its simplest prediction is that areas with larger submission growth show steeper AR declines. We find $r(\text{submission growth}_{2024\to2026},\,\Delta\text{AR})=-0.052$ ($p=0.86$). Datasets \& Benchmarks grew by $459\%$ while its AR changed by $-3.6$pp. This null result does not rule out changes in submission quality, including better entrants into growing areas; it only rejects the stated monotone growth--decline prediction. Figure~\ref{fig:robustness} (Appendix~\ref{app:robustness}) summarizes the diagnostics.\footnote{A promotional-language diagnostic is also null: title rhetoric measured with the 139-word lexicon from \citet{millar2022hype} is uncorrelated with topic-level AR ($r\approx0.01$, $p=0.94$); see Appendix~\ref{app:rhetoric}. This does not exclude every form of rhetorical influence.}

%% ============================================================
\section{A Measurement-Design Interpretation: Topic-Specific Reviewer Pools}
\label{sec:mechanism}
%% ============================================================

\subsection{Why Cross-Pool Comparability Must Be Designed}

Reviewer assignment is topic-structured: different topics are evaluated by different, only partly overlapping reviewer pools. We do not observe every aspect of pool composition and do not estimate the causal effect of reviewer assignment. The structural point is narrower: scores generated by different pools are not automatically equivalent merely because they share numerical labels. In the absence of designed empirical equating, the observed topic-specific score-to-decision mappings require either local-scale calibration or information beyond the score. Both interpretations make the recorded score incomplete as a venue-wide instrument.

Three dimensions may contribute to cross-pool differences:

\textbf{Expertise depth.} Topic-specific pools can contain different mixtures of reviewer experience, and prior reviewing history can affect how informative a numerical score is~\citep{stelmakh2021prior,shah2022challenges,tan2021calibration}. Our confidence diagnostic does not identify these mixtures. Expertise depth therefore remains a possible mechanism with unresolved causal status.

\textbf{Scoring culture.} Different research communities may develop different implicit interpretations of the rating scale~\citep{rogers2020improve}. Because reviewers are matched largely within communities, local scale use can aggregate into area-level offsets~\citep{tan2021calibration}. Such offsets need not be illegitimate, but a venue-wide score requires an explicit link among them.

\textbf{Novelty baseline.} Reviewers calibrate novelty against the work they see, so rapidly moving areas such as agent memory~\citep{xu2026contextual} and mature areas may use different comparison sets~\citep{wang2017bias,newman2009first}. Contribution scores exceed soundness scores across the 21 primary areas in our auxiliary data ($+14.5$pp first-mover association, $p=0.037$; promotional-word frequency grows $+0.77\%$/year, $p<0.001$). These associations motivate, but do not identify, area-dependent novelty baselines~\citep{millar2022hype}.

\subsection{How AC Discretion Handles an Incomplete Score}

When a score is incomplete, area chairs may rationally interpret it using written reviews, local scoring culture, perceived expertise, and other information~\citep{shah2022challenges,kargaran2025iclr}. Expert discretion can be a feature of peer review. The design problem is that the quantitative record does not show when a topic-specific adjustment reflects principled judgment, local-scale calibration, or unexplained drift.

Table~\ref{tab:topic_trends} makes temporal variation visible. Graph/Geometry Learning's acceptance rate relative to the venue average shifted from $+4.9$pp in 2024 to $-5.9$pp in 2026, a $-10.8$pp swing in two years. The data do not measure whether absolute submission quality changed, so this comparison is descriptive. Similarly, fitted thresholds dropped by 0.33--1.09 points across the 14 primary areas with three-year data while ICLR submissions grew from 7,237 to 18,790~\citep{papercopilot2024}. Reviewer-pool growth, submission composition, and changing decision policy are all possible explanations~\citep{beygelzimer2023neurips,kargaran2025iclr}.

The empirical result concerns the incompleteness of the recorded score as a venue-wide account of the decision. It does not evaluate whether AC discretion is correct. Without calibration information and explicit area criteria, an external audit cannot determine which topic effects reflect valid adjustment.

\textbf{The hype-cycle AR premium.}
Table~\ref{tab:topic_trends} isolates this dynamic more directly using topic-level data spanning all 6 years.
We classify 322 BERTopic-derived research topics into two groups: \emph{emerging} topics ($n_{23}\!\leq\!5$, growing to $n_{25}\!\geq\!30$ and $n_{26}\!\geq\!50$) and \emph{established-but-declining} topics (more than five papers by 2021, now growing slowly or shrinking). One paper can belong to multiple topics.

The contrast is stark.
The top-10 emerging topics (e.g., test-time scaling, masked diffusion language models, LLM-as-judge) carry an average acceptance rate of \textbf{32.5\%} in 2026 and \textbf{40.5\%} in 2025, well above the venue average of 28.5\% and 32.2\% respectively.
The top-10 established-but-declining topics (e.g., transfer learning, GANs, graph neural networks, CNNs) carry an average acceptance rate of only \textbf{24.6\%} in 2026, \textbf{3.9 percentage points below} the venue average.
The gap between the two groups is \textbf{7.9 pp} in 2026 and \textbf{16.0 pp} in 2025.
In 2025, emerging topics have both a higher median reviewer score (5.46 vs.\ 4.97) and a higher acceptance rate, so that comparison is not independent evidence of score--decision decoupling. In 2026, the median scores are close (4.29 vs.\ 4.22) while the acceptance-rate gap is 7.9\,pp. This group comparison remains descriptive; the joint conditional models and narrow score bands in Section~\ref{sec:inversion} carry the inferential claim.

The emerging-topic AR falls by 8.0pp from 2025 to 2026, compared with a 3.7pp venue-wide drop and a 2.9pp drop among the selected declining topics. This is consistent with a changing early-phase bar, but it could also reflect changes in submission composition or score use. We therefore present the trend as a hypothesis-generating pattern. It does not identify the cause of the initial difference~\citep{downs1972up,chu2021slowed}.

\begin{table}[t]
\setlength{\aboverulesep}{0.1ex}\setlength{\belowrulesep}{0.1ex}
\centering
\begin{small}
\caption{\textbf{Hype-cycle AR premium: emerging vs.\ established-but-declining topics (ICLR 2021--2026).}
Med.\ Score = median of per-paper mean reviewer ratings; Growth = $(n_{26}-n_{25})/n_{25}$.}
\label{tab:topic_trends}
\smallskip
\begin{tabular}{llrrrrrr}
\toprule
& \textbf{Topic} & $n_{25}$ & $n_{26}$ & \textbf{Growth} & \textbf{AR25} & \textbf{AR26} & \textbf{Med.\ Score} \\
\midrule
\multirow{10}{*}{\rotatebox[origin=c]{90}{\textbf{Top-10 Emerging}}}
& Test-time scaling & 34 & 323 & $+850\%$ & 55.9\% & 37.5\% & 5.71 / 4.50 \\
& Multimodal math reasoning & 30 & 178 & $+493\%$ & 40.0\% & 34.8\% & 5.42 / 4.50 \\
& MLLM complex reasoning & 100 & 390 & $+290\%$ & 33.0\% & 31.8\% & 5.25 / 4.50 \\
& Mathematical reasoning & 85 & 327 & $+285\%$ & 38.8\% & 31.2\% & 5.50 / 4.40 \\
& LLM-as-judge & 72 & 230 & $+219\%$ & 41.7\% & 29.1\% & 5.50 / 4.00 \\
& Flow matching & 85 & 257 & $+202\%$ & 40.0\% & 37.0\% & 5.50 / 4.50 \\
& LLM agents & 107 & 318 & $+197\%$ & 29.9\% & 26.1\% & 5.25 / 4.00 \\
& Hallucination mitigation & 39 & 108 & $+177\%$ & 41.0\% & 28.7\% & 5.50 / 4.00 \\
& Masked diffusion LM & 116 & 306 & $+164\%$ & 41.4\% & 37.3\% & 5.29 / 4.50 \\
& Tokenizer/tokenization & 55 & 133 & $+142\%$ & 43.6\% & 31.6\% & 5.67 / 4.00 \\
\cmidrule{2-8}
& \textit{Group average} & & & & \textit{40.5\%} & \textit{32.5\%} & \textit{5.46 / 4.29} \\
\midrule
\multirow{10}{*}{\rotatebox[origin=c]{90}{\textbf{Top-10 Declining}}}
& Transfer learning & 92 & 76 & $-17\%$ & 25.0\% & 22.4\% & 4.90 / 4.00 \\
& SAM / segmentation & 78 & 69 & $-12\%$ & 23.1\% & 20.3\% & 4.88 / 4.50 \\
& Activation functions & 79 & 74 & $-6\%$ & 31.6\% & 33.8\% & 5.00 / 4.50 \\
& CNN / convolution & 165 & 168 & $+2\%$ & 20.6\% & 16.1\% & 4.60 / 4.00 \\
& Semantic segmentation & 94 & 96 & $+2\%$ & 34.0\% & 25.0\% & 5.25 / 4.00 \\
& GANs & 76 & 78 & $+3\%$ & 23.7\% & 25.6\% & 4.58 / 4.00 \\
& SGD / gradient descent & 261 & 275 & $+5\%$ & 36.4\% & 34.9\% & 5.25 / 4.50 \\
& Graph neural networks & 416 & 444 & $+7\%$ & 28.6\% & 19.8\% & 5.00 / 4.00 \\
& Active learning & 56 & 60 & $+7\%$ & 17.9\% & 15.0\% & 4.71 / 4.00 \\
& Bandits & 70 & 76 & $+9\%$ & 34.3\% & 32.9\% & 5.50 / 4.67 \\
\cmidrule{2-8}
& \textit{Group average} & & & & \textit{27.5\%} & \textit{24.6\%} & \textit{4.97 / 4.22} \\
\bottomrule
\end{tabular}
\end{small}
\vspace{-0.5cm}
\end{table}

\subsection{A Common Rubric Still Requires Empirical Equating}

A common rubric and reviewer training can reduce ambiguity, but common wording alone does not demonstrate that the scale is empirically equivalent across pools. A score of 6 may compress domain-specific tradeoffs such as stealth and potency in backdoor research~\citep{xu2026breaking}, while also meaning ``high relative standing in this area'' or ``meets a venue-wide standard.'' When local comparison sets and criteria differ, these interpretations require a designed link. Shared anchor papers, overlapping calibrated raters, or post-hoc equating could in principle provide such a link. The current public process documents none of them.

This is not a claim about reviewer effort. It is a measurement-design problem: the instrument is used for a cross-pool comparison without evidence of cross-pool equating~\citep{shah2022challenges,tan2021calibration}.

A related objection holds that ACs may rationally trade off the \emph{recorded} Contribution and Soundness subscores. Under a simple tradeoff on those recorded scales, their decision boundaries should move in opposite directions. Both estimated thresholds rise in the high-threshold group (Soundness 3.48 vs.\ 3.31; Contribution 3.52 vs.\ 3.29; both $p<0.05$; Appendix~\ref{app:subscore}). This rejects that simple recorded-subscore tradeoff. It does not establish that subscores are comparable across areas, nor does it exclude area-specific meanings or valid information absent from the subscores. The main conditional result does not rely on this diagnostic.

%% ============================================================
\section{Alternative Views}
\label{sec:altviews}
%% ============================================================

We address three genuinely held positions that are opposed to our thesis.

\textbf{View 1: Topic stratification reflects legitimate community norms.}
One can argue that different research communities within ML operate with genuinely different standards for novelty, rigor, and impact~\citep{kuhn1962structure}. A theoretical ML paper and a data-free online defense paper~\citep{xu2026internal} cannot be evaluated on the same scale; asking them to have the same acceptance rate imposes an artificial uniformity on a field that is legitimately diverse. On this view, the ``bias'' we document is not bias at all. It is the appropriate expression of pluralistic scientific judgment~\citep{merton1973sociology}.

This view is partly correct: different local score meanings may be legitimate, and our data do not support imposing uniform acceptance rates. Nor should a same-score gap disappear when the same number has different local meanings. The measurement-design failure arises when those local scores serve as one nominally common instrument without an explicit, auditable calibration. The same-score-band analysis (Figure~\ref{fig:samescorear}) establishes that a topic-specific decoder is empirically necessary; conditional reporting would let researchers, area chairs, and program committees examine whether the resulting disparities reflect stated scientific standards.

The ACL 2023 experience also shows that process design can change area patterns~\citep{rogers2020improve}. After deliberate matching between reviewers and contribution types, Rogers et al.~\citep{rogers2023acl} report that the earlier disadvantage for non-mainstream contribution types did not persist; their Table~7 shows a remaining difference in favor of non-mainstream areas, not zero difference. We treat this as evidence that measurement and assignment policy matter, not as proof that all area differences are artifacts.

\textbf{View 2: Structural solutions risk gaming and unintended consequences.}
Any structural intervention (normalized scoring, quotas, calibration papers) creates new gaming opportunities~\citep{balietti2016peer}. If researchers learn that Foundation/LLMs is a soft landing, they will relabel marginal work as foundation model research. Quotas that floor the acceptance rate in cold areas may accept papers that the relevant community genuinely believes are below-threshold~\citep{stelmakh2021peerreview4all}. These are real concerns, and we do not claim our proposed solutions are costless.

Our response is that the \emph{status quo also produces gaming}~\citep{lipton2019troubling,abdalla2023elephant}: researchers already optimize their area labels to improve acceptance probability. The question is whether transparent, designed incentives are better or worse than opaque, emergent ones. We believe transparency (publishing score-conditional acceptance rates by area) has low cost and creates the accountability needed for the community to self-correct.

\textbf{View 3: Mean reviewer score is not a neutral quality proxy; AC decisions may rationally correct for its limitations.}
One can argue that reviewer scores are a composite of Soundness and Contribution whose relative weights legitimately differ by area. An AC who upweights Contribution for a frontier area and Soundness for a mature area may be making a principled judgment. Under this view, the score-AR inversion reflects a correction, and an intervention that constrains AC discretion could reduce review quality.

We take this objection seriously. The subscore diagnostic in Section~\ref{sec:mechanism} does not support a simple tradeoff on the two recorded subscore scales. Both estimated thresholds move upward, with no opposite movement across the two dimensions. However, areas may interpret those subscores differently, and ACs may use valid information that they do not record. We therefore do not claim to falsify rational AC judgment in general. This view reinforces our design claim: if the score and recorded subscores are locally interpreted or incomplete, the adjustment should be made explicit enough to audit.

%% ============================================================
\section{Call to Action}
\label{sec:solutions}
%% ============================================================

The measurement-design problem is remediable. In one borderline-zone comparison (aligned score 4--6), the observed acceptance rates are 19.0\% for graph contrastive learning and 53.8\% for object-centric learning, a 2.8$\times$ gap that the recorded score alone does not encode. This does not establish which decisions are substantively correct. It shows why the mapping from local review evidence to the venue-wide decision should be explicit. We propose three interconnected actions. The first two act directly at the area level; the third supplies partial individual-level safeguards.

\subsection{Action 1 (Program Committees): Publish Area-Stratified Acceptance Data}

The minimum intervention is also the most powerful: after each review cycle, publish score-conditional acceptance rates stratified by primary area, with additional breakdowns by Soundness and Contribution subscores separately. This requires only a dashboard, not a policy change. ICLR's existing full-review disclosure policy makes it immediately feasible.

Transparency serves two functions. First, it creates accountability: once conditional outcomes are reported, the community can track whether disparities persist and whether interventions change them. Second, it sharpens the normative dispute in View~3. Persistent gaps after conditioning on recorded scores and subscores would show that the recorded instruments remain incomplete; disappearing gaps would show that those instruments account for the observed mapping. Neither pattern alone proves whether latent-quality differences or unrecorded judgments are justified, but the report makes assumptions and changes visible. Opacity leaves these questions untestable; transparency makes them empirically tractable.

Beyond ICLR, we call on venues to treat area-stratified, score-conditional acceptance rates as a first-class audit metric, alongside gender and institutional-origin statistics~\citep{tomkins2017reviewer,lee2013bias}. The ACL experience is instructive: Rogers et al.~\citep{rogers2023acl,rogers2020improve} documented an earlier disadvantage for non-mainstream contribution types and a later reversal after deliberate reviewer matching. Measurement preceded change. Whether analogous patterns occur at NeurIPS or ICML is unknown and requires representative data.

\subsection{Action 2 (Research Communities): Develop Area-Specific Review Rubrics}

Research areas legitimately emphasize different evidence: a theory paper may be judged by proofs and generality, a systems paper by reproducibility and engineering rigor, a security defense by poisoned-data separation~\citep{xu2025clip}, and an empirical paper by experimental breadth and ablations. Compressing these judgments into one 1--10 score without recording area-specific criteria makes the basis for cross-area interpretation unclear.

We call on each primary research community within ML to develop and maintain \textit{area-specific review rubrics}: publicly endorsed checklists of what the community regards as load-bearing criteria for acceptance~\citep{rogers2020improve,wang2025criteria}. This is not a call for bureaucratic standardization. The rubric for reinforcement learning should look different from the rubric for probabilistic methods, while autonomous-agent risk may require trace-economic evidence~\citep{xu2026agent}. The goal is to make implicit community standards explicit, providing a shared reference point that reduces the most common calibration failures without constraining expert judgment.

This intervention is increasingly relevant as submission and reviewer pools grow~\citep{ma2025acl,papercopilot2024,kim2025crisis,azad2024trends}. Newer reviewers may lack the tacit knowledge that experienced reviewers use to interpret ``above the bar'' within an area~\citep{tran2020open,stelmakh2021novice,kargaran2025iclr,kim2025crisis}. Area-specific rubrics can make that knowledge explicit while preserving pluralism. They will not by themselves equate numerical scales, but they make the criteria behind local judgments inspectable.

\subsection{Action 3 (Venues): Supplement Scores with Calibrated Signals}

Transparency and area-specific criteria are the primary area-level remedies. The following mechanisms operate mainly at the individual level and should be treated as complementary safeguards, not complete solutions to cross-area drift.

\textit{Reviewer calibration profiling}: a reviewer's score distribution can encode personal scoring style as well as paper mix. With sufficient history and appropriate adjustment for assignment, a calibration profile may help ACs interpret unusually strict or lenient raters~\citep{tan2021calibration}. This reduces one upstream source of cross-area drift when reviewer mixtures differ, but it cannot align area criteria by itself.

\textit{Author self-ranking via the Isotonic Mechanism}~\citep{su2022isotonic,su2024icml}: authors with multiple submissions rank their own papers, and the mechanism projects raw scores to be consistent with this ordering. Its ICML 2023 experiment reported a reduction in score MSE ($p<0.01$), and it was adopted as ICML 2026 policy. This supplies a limited within-author cross-area signal; it applies only to multi-submission authors and cannot resolve venue-wide area drift on its own.

\textit{Foregrounding written reviews at the borderline}: numerical scores are a lossy compression of a reviewer's judgment. A score of 6 paired with ``technically sound, clearly written, incremental contribution'' and a score of 6 paired with ``potentially transformative idea, execution needs work'' are not equivalent, but mean-score aggregation treats them identically. We advocate for a norm in which area chairs engage with the substance of written reviews and document which claims were decisive in borderline decisions, including agent work whose contribution hinges on context management~\citep{xu2026llm,kang2018dataset,liang2024llm}. This practice is already informal in the best decisions; making it the expected norm reduces the degree to which miscalibrated scores from reviewers without established profiles determine outcomes.

Together, the actions form a layered program: conditional reporting reveals area-level mappings, explicit rubrics record area criteria, and individual calibration signals reduce one source of upstream variation. The first two address the level at which we diagnose the pattern; the third is supplementary.

%% ============================================================
\section{Related Work}
\label{sec:related}
%% ============================================================

Peer review inconsistency is well-documented. The NeurIPS 2014 experiment~\citep{cortes2021inconsistency} found 50\% of score variance was subjective; the 2021 replication~\citep{beygelzimer2023neurips} showed arbitrariness growing with scale; Goldberg et al.~\citep{goldberg2025review} conducted the first RCT on review quality, finding inconsistency even among trained reviewers. Reviewer biases compound this inconsistency: single-blind review introduces institutional prestige and gender biases~\citep{tomkins2017reviewer,tran2020open,lee2013bias}, and novice reviewers penalize resubmissions~\citep{stelmakh2021prior}. Tan et al.~\citep{tan2021calibration} and Su~\citep{su2022isotonic,su2024icml} propose score calibration to correct idiosyncratic scales. Our work adds an area-level audit and a measurement-design interpretation.

Topic-stratified analysis of ML peer review is nascent. Rogers et al.~\citep{rogers2023acl,rogers2020improve} study reviewer--contribution matching at NLP venues; Jung et al.~\citep{jung2025process} analyze process-centric signals in ICLR 2017--2025. To our knowledge, our work is the first large-scale audit of topic-level score--acceptance mapping with direct same-score comparisons and a joint conditional test. Hype cycles~\citep{downs1972up} and bias against novelty~\citep{wang2017bias,chu2021slowed} motivate hypotheses about emerging topics, but our observational trends do not identify those mechanisms. Prior work on reviewer bias focuses mainly on individual or institutional effects~\citep{tomkins2017reviewer,abdalla2023elephant,sun2021doubleblind}. Our question concerns whether a common recorded score has a stable area-level decision meaning. Kang et al.~\citep{kang2018dataset} introduced PeerRead for computational review analysis; we extend this line using ICLR's full-review disclosure policy.

%% ============================================================
\section{Conclusion}
\label{sec:conclusion}
%% ============================================================

Across 50{,}289 ICLR papers (2021--2026), topic rates span 18.3\% to 46.0\% among the 146 qualifying overlapping topics, and a joint topic block improves acceptance prediction beyond recorded score and year ($\chi^2=341.2$, $p=5.1\times10^{-3}$; paper-clustered $p=1.8\times10^{-2}$). Direct narrow-band rates also vary at every reported score level; 8$\times$ is the observed maximum, while the 2026 score-$\approx5.0$ median pairwise gap is 6.18pp. These results show that a score has no stable observed decision meaning across topics. They do not imply equal latent quality at equal scores or that every area adjustment is wrong.

We identify the lack of explicit cross-pool calibration as a \textbf{measurement-design failure}. The primary remedies are area-stratified, score-conditional reporting and explicit area criteria; reviewer profiles~\citep{tan2021calibration} and author self-rankings~\citep{su2022isotonic,su2024icml} are partial supplements. Our empirical conclusion is confined to ICLR. Whether the pattern generalizes to NeurIPS or ICML is unknown, and representative data are required to answer that question. Transparency is the prerequisite for making local calibration and discretionary judgment auditable.

%% ============================================================
%% BIBLIOGRAPHY
%% ============================================================

\bibliographystyle{plainnat}
\bibliography{references}

%% ============================================================
%% APPENDIX
%% ============================================================

\appendix

\newpage

%% ============================================================
\section{Main Results Figures}
\label{app:figures}
\label{app:systematic}
%% ============================================================

\begin{figure}[H]
\centering
\vspace{-0.3cm}
\includegraphics[width=1.0\linewidth]{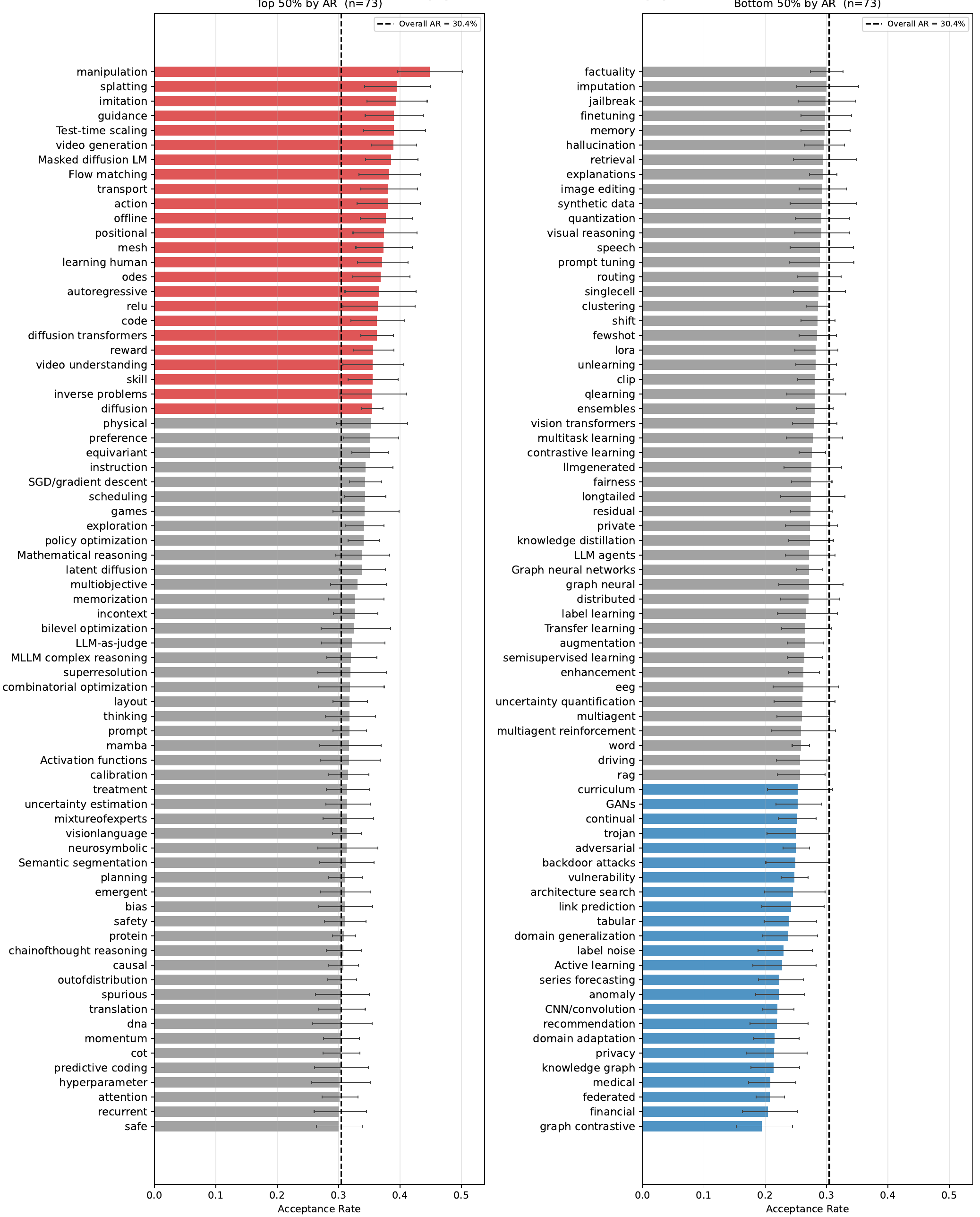}
\vspace{-0.5cm}
\caption{\textbf{Acceptance rate disparities across 146 qualifying BERTopic topics at ICLR 2021--2026.}
Each bar shows the acceptance rate (with 95\% Wilson CI) for one topic. Left panel: top 50\% of topics by AR (highest AR first). Right panel: bottom 50\% of topics by AR (lowest AR last). Only topics with $\geq$250 total matched papers and $\geq$30 borderline papers are included. Dashed horizontal line indicates the pooled average (30.4\%). Red bars: $>$5pp above average; blue bars: $>$5pp below average; colors do not encode statistical significance. Across the 146 topic-versus-pooled tests, 49 survive BH, 31 BY, and 19 Holm correction at 0.05 (Appendix~\ref{app:multiplicity}).}
\vspace{-0.1cm}
\label{fig:systematic}
\end{figure}

\begin{figure}[H]
\centering
\vspace{-0.2cm}
\includegraphics[width=\linewidth]{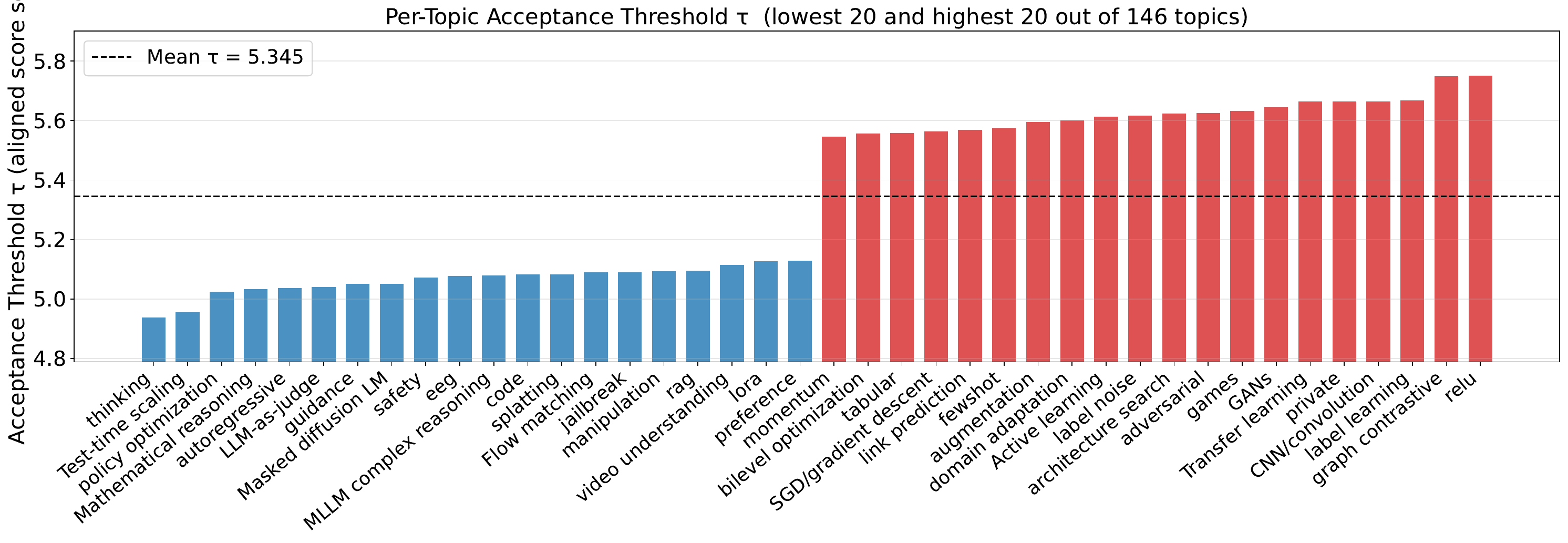}
\vspace{-0.7cm}
\caption{\textbf{Per-topic acceptance threshold analysis (ICLR 2021--2026).}
Each bar shows $\hat{\tau}$, the mean reviewer score at which a paper has a fitted 50\% acceptance probability (estimated by per-topic logistic regression on within-year normalized scores). Blue: 20 lowest-threshold topics; red: 20 highest-threshold topics, out of 278 qualifying topics. Spread: 0.812 points (chain-of-thought: $\hat{\tau}=4.938$; activation functions: $\hat{\tau}=5.751$). The threshold is a descriptive summary, not a latent-quality model; the narrow-band analysis does not use it.}
\vspace{-0.3cm}
\label{fig:threshold}
\end{figure}

\begin{figure}[H]
\centering
\vspace{-0.3cm}
\includegraphics[width=\linewidth]{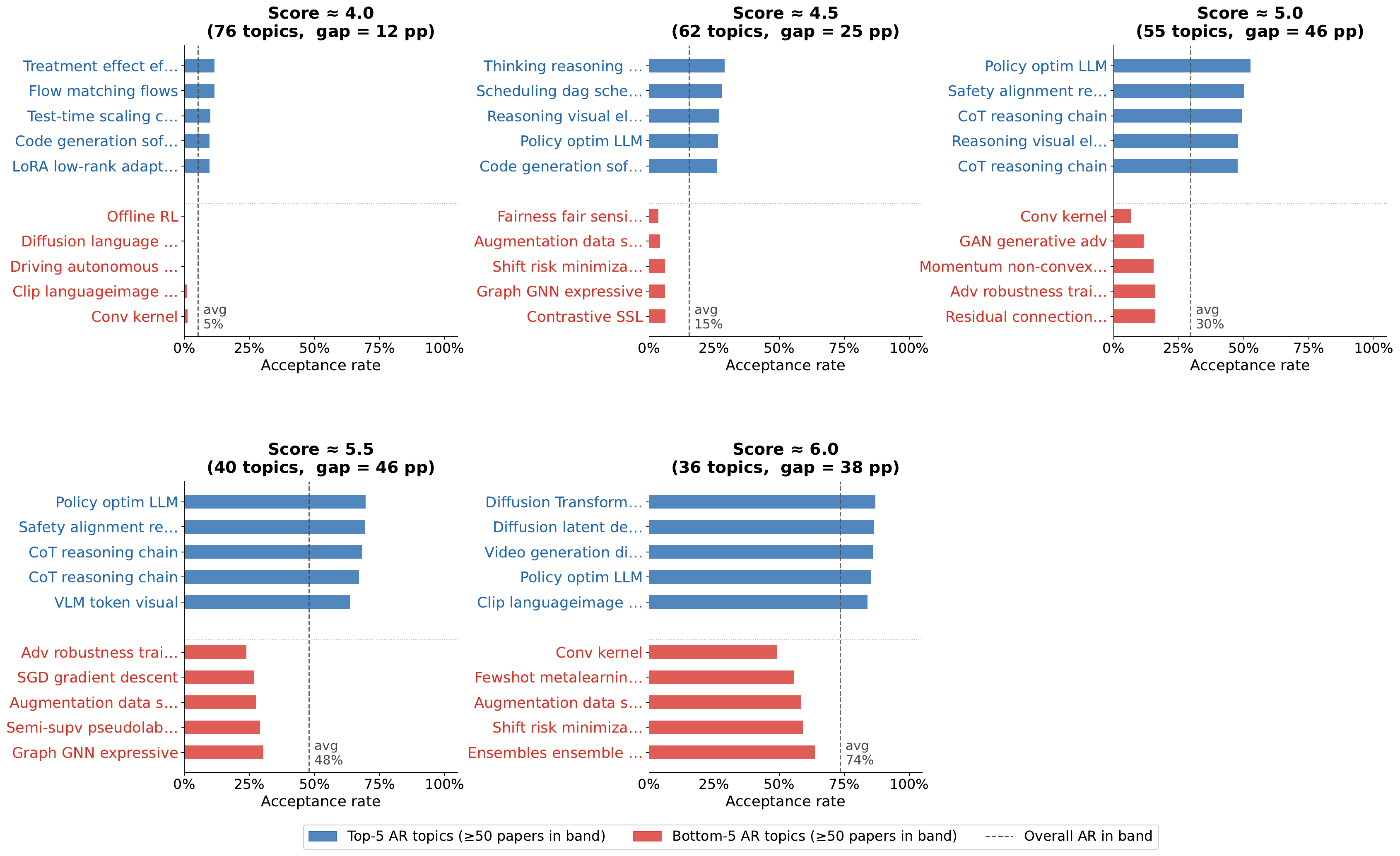}
\vspace{-0.3cm}
\caption{\textbf{At identical recorded reviewer scores, the observed maximum ratio reaches 8$\times$ across topics (ICLR 2021--2026).}
For each of five score bands ($\pm0.125$ around 4.0, 4.5, 5.0, 5.5, 6.0), the top-5 (blue) and bottom-5 (red) topics by acceptance rate are shown, among all topics with $\geq$50 papers in the band. The dashed line marks the overall acceptance rate in each band. At score $\approx$4.5 and $\approx$5.0 the displayed ratio reaches 8$\times$. This is an extreme descriptive contrast, not the typical effect; in 2026 at score $\approx5.0$, the median pairwise gap is 6.18pp and the P90--P10 spread is 15.29pp.}
\vspace{-0.3cm}
\label{fig:samescorear}
\end{figure}

\textbf{Technical note} (Figure~\ref{fig:systematic})\textbf{.}
Acceptance rates are computed as accepted papers divided by effective submissions (accepted + rejected, excluding desk-reject-equivalent withdrawals defined as papers withdrawn before or during early reviewing with fewer than 3 confidence scores). We use 95\% Wilson confidence intervals~\citep{wilson1927probable} because they have better coverage than Wald intervals when acceptance rates are near 0 or 1, which occurs for several small-to-medium topics. The dual-panel layout prevents overlapping labels; within each panel, topics are sorted by descending acceptance rate. The inclusion filter ($\geq$250 matched papers across all six years and $\geq$30 borderline papers with aligned score $\in[4,6]$) ensures that both the overall AR estimate and the logistic threshold estimate are based on sufficient data. Topics failing this filter are reported in Appendix~\ref{app:topic_stats} but excluded from threshold and same-score-band analyzes. The pooled average (30.4\%) is computed over all 50,289 effective submissions and serves as the reference line; it does not equal the unweighted mean of topic ARs because topics vary greatly in size.

\subsection{Per-Topic Acceptance Threshold Analysis}
\label{app:threshold}

\textbf{Technical note.}
The acceptance threshold $\hat{\tau}$ is derived from a per-topic logistic regression:
\begin{equation}
P(\text{accept} \mid \text{score}) = \sigma(\beta_0 + \beta_1 \cdot \text{score}),
\end{equation}
yielding $\hat{\tau} = -\hat{\beta}_0/\hat{\beta}_1$, i.e., the aligned mean reviewer score at which a paper has a fitted 50\% probability of acceptance. Only fits satisfying $\hat{\beta}_1 > 0.3$ (positive discrimination) and $\hat{\tau} \in [3.0, 8.5]$ are retained, leaving 278 qualifying topics out of 322. The aligned score is the within-year standardized reviewer mean (z-scored to zero mean and unit variance across all papers in that year), then rescaled to the original 1--10 range, to remove year-level threshold drift. Figure~\ref{fig:threshold} displays the 20 topics with the lowest and highest estimated thresholds. Error bars show the 95\% bootstrap confidence interval for $\hat{\tau}$ (500 resamples). Mean confidence is not positively associated with threshold ($r=-0.102$, $p=0.626$), which fails to support a simple recorded-confidence explanation but does not establish equivalent reviewer expertise.

\subsection{Same-Score-Band Acceptance Rate Comparison}
\label{app:samescorear}

\begin{figure}[t]
\centering
\includegraphics[width=\linewidth]{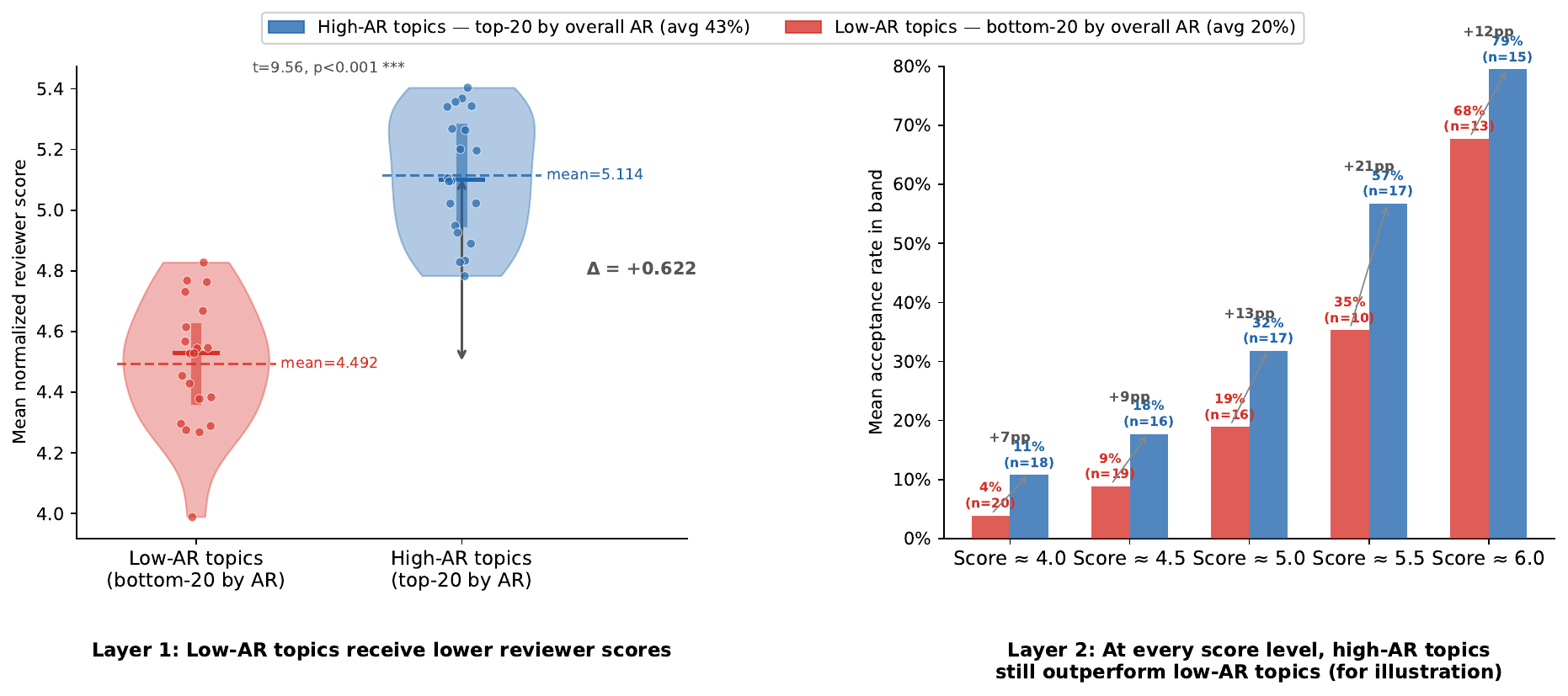}
\caption{\textbf{Illustrative mechanism: topics with low overall AR face a double disadvantage (ICLR 2021--2026).}
\emph{Left}: Distribution of mean normalized reviewer scores for the top-20 and bottom-20 AR topics ($n\geq50$, grouped by overall six-year AR). Low-AR topics receive scores 0.622 points lower on average ($t=9.56$, $p<0.001$). \emph{Right}: At each score band, the mean acceptance rate of the two groups. The high-AR group has higher acceptance rates at every score level; the gap peaks at score $\approx$5.5 (+21\,pp). Note: the grouping is by construction correlated with AR and is offered as a mechanistic illustration only. The primary nonparametric evidence is Table~\ref{tab:samescore}, which requires no topic grouping.}
\label{fig:doubledisadvantage}
\end{figure}

\textbf{Technical note.}
Figure~\ref{fig:samescorear} is the primary nonparametric evidence for score--acceptance decoupling. For each center score $s \in \{4.0, 4.5, 5.0, 5.5, 6.0\}$, we select all papers whose mean reviewer rating (unaligned, original 1--10 scale) falls in the window $[s-0.125,\, s+0.125]$. For each topic, we compute the acceptance rate among all papers in that window belonging to that topic. Topics with fewer than 50 papers in the window are excluded to ensure stable AR estimates (the 50-paper threshold was chosen to keep the half-width of the 95\% Wilson CI below 14 percentage points). The top-5 and bottom-5 topics by in-band AR are then displayed. No logistic fitting, z-scoring, or parametric decision curve is involved. The choice of a $\pm0.125$ window trades precision of ``same score'' against sample size; sensitivity checks with $\pm0.25$ and $\pm0.0625$ windows are qualitatively similar. The ratio is undefined at score 4.0 because the displayed bottom-5 AR is 0.0\%; Table~\ref{tab:samescore} reports N/A. Because selecting extremes inflates their apparent separation, we label 8$\times$ as the observed maximum. As a less selection-sensitive summary, the 2026-only score-$\approx5.0$ comparison contains 47 qualifying topics, with a median absolute pairwise gap of 6.18pp and a P90--P10 spread of 15.29pp.

\subsection{Double Disadvantage Mechanism}
\label{app:doubledisadvantage}

\textbf{Technical note.}
Figure~\ref{fig:doubledisadvantage} uses a \emph{split-by-AR-group} design that introduces a circularity by construction: topics are sorted by six-year AR to form groups, and then in-band ARs (which contribute to the six-year AR) are plotted. The figure is therefore offered only as a mechanistic illustration, not independent evidence. Its purpose is to decompose the score-acceptance gap into two additive components (reviewer score offset and within-score-band AR gap) for didactic clarity. The grouping uses the top-20 and bottom-20 topics by overall six-year acceptance rate among 219 topics with $n\geq50$ matched papers; the $n\geq50$ filter ensures stable AR estimates. Left panel: kernel density estimates of the distribution of within-year z-scored reviewer means for all papers belonging to each group. The score difference ($\Delta=0.622$, $t=9.56$, $p<0.001$) is computed via a two-sample $t$-test on paper-level scores, clustering standard errors at the topic level. Right panel: for each score band (same $\pm0.125$ windows as Figure~\ref{fig:samescorear}), the mean AR across topics in each group is plotted; shaded bands are $\pm1$ SE across topics. The monotone ordering (high-AR group always above low-AR group) holds at every score level and is the key takeaway, independent of the circularity caveat.

\section{Logistic Regression Methodology}
\label{app:logit}

\subsection*{Score Field Normalization Across Years}

ICLR changed its review score field names between cohorts. The unified \texttt{Paper.rating} attribute used in all analyzes is constructed as follows:
\begin{itemize}
\item \textbf{2021, 2024--2026}: sourced from \texttt{rating\_avg} (a float representing the mean of all reviewer ratings for that paper).
\item \textbf{2022--2023}: sourced from \texttt{recommendation\_avg}, which stores a \texttt{[mean, std]} list; the mean component is extracted.
\end{itemize}
In all years the rating scale is nominally 1--10. Papers for which neither field is populated (i.e., papers that were withdrawn before any reviewer submitted a score, or desk-rejected submissions) are assigned \texttt{rating=None} and excluded from all score-conditional analyzes. The number of reviewers per paper (\texttt{n\_reviewers}) is used to verify that the mean score is based on at least one completed review; the standard deviation (\texttt{rating\_std}) is available as a within-paper dispersion measure but is not used in the main analyzes.

\subsection*{Per-Topic Logistic Threshold Estimation}

For each topic with at least 80 papers, we fit the logistic model:
\begin{equation}
P(\text{accept} \mid \text{score}) = \sigma(\beta_0 + \beta_1 \cdot \text{score})
\end{equation}
where $\sigma(z) = (1+e^{-z})^{-1}$ is the sigmoid function and \texttt{score} is the within-year z-scored mean reviewer rating rescaled to the original 1--10 range (``aligned score''). Parameters are estimated by maximum likelihood using the L-BFGS-B optimizer with analytic gradients and a tolerance of $10^{-8}$ on the log-likelihood. Standard errors are computed from the observed Fisher information matrix. The acceptance threshold is $\hat{\tau} = -\hat{\beta}_0/\hat{\beta}_1$, the score at which predicted acceptance probability equals 50\%.

We apply two post-fit quality filters:
\begin{itemize}
\item $\hat{\beta}_1 > 0.3$: ensures the fitted curve exhibits positive discrimination (higher scores correlate with higher acceptance probability). Topics where this fails typically have extremely low or extremely high baseline AR, making the logistic slope unreliable.
\item $\hat{\tau} \in [3.0, 8.5]$: excludes threshold estimates that lie far outside the observed score range, indicating poor model fit or insufficient data near the decision boundary.
\end{itemize}
Of 322 BERTopic topics, 278 pass both filters and are used in threshold analyzes.

\subsection*{Multi-Level Logistic Regression for the Likelihood Ratio Test}

To test whether topic identity explains acceptance variance beyond reviewer scores, we fit two nested logistic models on the full dataset of 113,290 paper-topic pairs (44,177 unique papers, one row per paper-topic match):

\begin{align}
\text{Model 0 (baseline):} &\quad \text{logit}\,P(\text{accept}) = \alpha + \beta\cdot\text{score} + \gamma\cdot\text{year} \\
\text{Model 1 (topic):} &\quad \text{logit}\,P(\text{accept}) = \alpha + \beta\cdot\text{score} + \gamma\cdot\text{year} + \sum_{k=1}^{277}\delta_k\cdot\mathbf{1}[\text{topic}=k]
\end{align}

The likelihood ratio statistic $\Lambda = -2(\ell_0 - \ell_1)$ follows a $\chi^2$ distribution with 277 degrees of freedom under $H_0$ (topic dummies add no explanatory power). We obtain $\Lambda = 341.2$, $p = 5.1\times10^{-3}$. Because papers can belong to multiple topics, standard errors and $p$-values are validated with a cluster-robust sandwich estimator clustering at the paper level, yielding $p = 1.8\times10^{-2}$. Both specifications reject $H_0$ at conventional significance levels.

\subsection*{One-Row-per-Paper Replication}

The paper--topic-pair model duplicates papers with multiple topic memberships. We therefore run a stricter replication with one row for each of 48,800 decided papers with a recorded score. Let $Z_{ik}$ indicate membership in each of the 146 overlapping topics that meet the Figure~\ref{fig:systematic} filter. To allow the score--decision relationship to vary flexibly over time, the baseline contains a separate intercept and cubic score curve for every year:
\begin{align}
\text{Baseline:}\quad &\operatorname{logit}P(A_i)=\sum_y \mathbf{1}[Y_i=y]f_y(S_i), \\
\text{Topic model:}\quad &\operatorname{logit}P(A_i)=\sum_y \mathbf{1}[Y_i=y]f_y(S_i)+\sum_{k=1}^{146}\delta_k Z_{ik},
\end{align}
where each $f_y$ is cubic. All 146 topic indicators enter jointly. The likelihood-ratio statistic is $\chi^2=180.4$ with 146 degrees of freedom ($p=2.79\times10^{-2}$). This check avoids duplicated outcome rows and relaxes the linear score assumption; it supports the same conclusion that the topic block contains decision information beyond recorded score and year.

\subsection{Observable-Signal Adjustment}
\label{app:observable-adjustment}

A remaining explanation is that the recorded mean score omits other information used in the decision. We test the observable part of this explanation with one row for each of the 48,800 decided papers. The all-year signal set contains within-paper rating dispersion, mean and dispersion of reviewer confidence, reviewer count, and the mean and dispersion of review length. A second specification adds author and reviewer reply counts and reply lengths as interaction proxies. For 37,531 papers from 2024 to 2026, a third specification adds the mean and dispersion of Soundness, Contribution, and Presentation. Separate diagnostics use empirical and technical novelty subscores for 8,222 papers from 2022 and 2023.

\begin{figure}[t]
\centering
\includegraphics[width=\linewidth]{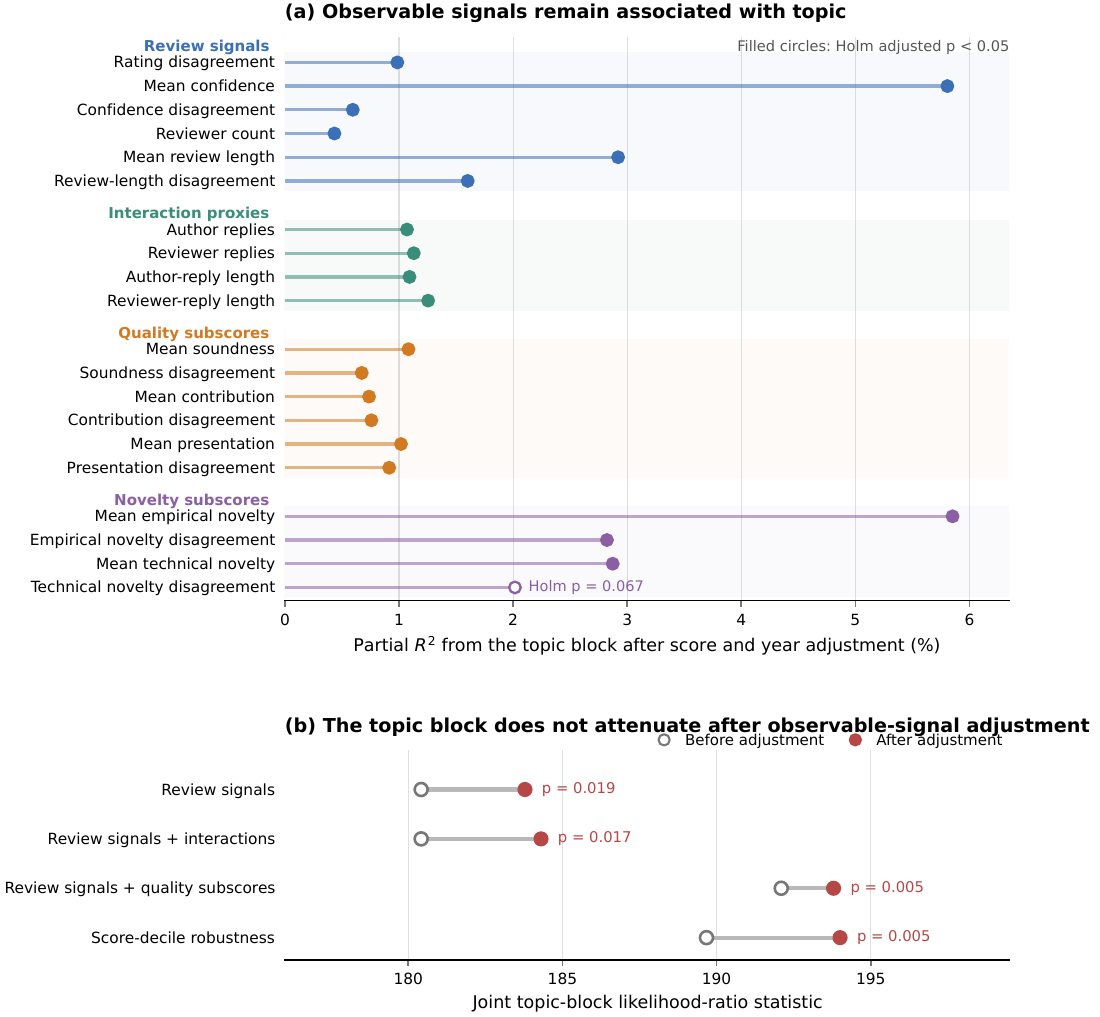}
\caption{\textbf{Observable-signal adjustment leaves the joint topic result intact.}
\emph{(a)} Partial $R^2$ from the joint topic block for each auxiliary signal after a year-specific cubic adjustment for mean score. Filled circles pass Holm correction within their signal family. Quality subscores are available from 2024 to 2026, and novelty subscores from 2022 and 2023. \emph{(b)} Joint likelihood-ratio statistic for the 146-topic block before and after adding the indicated auxiliary set. Each connected pair uses the same papers and score control. The final row replaces cubic score curves with year-by-score-decile fixed effects. The small increases in the statistic do not establish a stronger topic effect; the supported conclusion is that the observed auxiliary signals do not attenuate the joint topic result.}
\label{fig:observable-adjustment}
\end{figure}

We first test whether each auxiliary signal is conditionally independent of topic after controlling for mean score and year. For signal $X_j$, we fit a year-specific cubic score curve and jointly add the same 146 overlapping topic indicators used in the one-row replication. Figure~\ref{fig:observable-adjustment}a reports the topic block's partial $R^2$. After Holm correction within each prespecified signal family, 19 of 20 signals retain a significant topic association. Partial $R^2$ values range from 0.43\% to 5.85\%. Technical-novelty disagreement is the only exception ($p_{\mathrm{Holm}}=0.067$). Thus these signals are neither fully absorbed by the recorded score nor distributed identically across topics.

We next add the auxiliary signals to the acceptance model. Each signal receives separate linear and quadratic terms within each year. The auxiliary blocks themselves contain acceptance information beyond mean score and year: $\chi^2=234.2$ ($df=72$, $p=4.74\times10^{-19}$) for the six review signals, $\chi^2=377.0$ ($df=112$, $p=2.10\times10^{-30}$) after adding the interaction proxies, and $\chi^2=312.9$ ($df=72$, $p=8.98\times10^{-32}$) for the review signals plus the three quality subscores in 2024 to 2026.

The central result is that the joint topic test does not decrease after these adjustments. With the six review signals, the topic statistic changes from 180.42 ($p=0.0279$) to 183.79 ($p=0.0186$). Adding interaction proxies yields 184.31 ($p=0.0175$). In the 2024 to 2026 subscore sample, it changes from 192.10 ($p=0.0063$) to 193.80 ($p=0.0050$). Replacing the cubic score curves with year-by-score-decile fixed effects gives the same pattern, from 189.67 ($p=0.0087$) to 194.01 ($p=0.0048$). All reported logistic models converge.

This analysis rules out the measured variables as a sufficient explanation of the residual topic association. It does not rule out unobserved review or rebuttal semantics, ethics and fit judgments, or committee discussion. Interaction variables may also occur after the initial scores, so their coefficients have no causal interpretation. The residual dependence supports a topic-conditioned decision regime, not a causal effect of topic identity. Operationally, any cross-topic interpretation of the recorded score must therefore condition on an implicit or explicit model of topic-specific scoring and decision regimes. The data do not identify which actor applies that model or at which stage.

\subsection*{Individual-Topic Tests and Multiple-Testing Correction}
\label{app:multiplicity}

Figure~\ref{fig:systematic} includes 146 topic-versus-pooled acceptance-rate comparisons. We treat these as one family. For topic $k$, the submitted analysis uses the two-sided normal statistic
\begin{equation}
z_k=\frac{\widehat{\mathrm{AR}}_k-\widehat{\mathrm{AR}}_{\mathrm{pool}}}{\sqrt{\widehat{\mathrm{AR}}_{\mathrm{pool}}(1-\widehat{\mathrm{AR}}_{\mathrm{pool}})/n_k}}.
\end{equation}
There are 59 nominal $p<0.05$ results. At the 0.05 level, 49 remain under Benjamini--Hochberg FDR correction, 31 under Benjamini--Yekutieli FDR correction (valid under arbitrary dependence), and 19 under Holm family-wise correction. Topic overlap induces dependence, which is why we report both BH and the more conservative BY and Holm results. These corrections change which individual topics are labeled significant; they do not apply to the separate single joint tests of the complete topic block.

\subsection*{Recorded-Subscore Tradeoff Diagnostic}
\label{app:subscore}

The subscore diagnostic repeats the per-topic logistic procedure with \texttt{soundness} and \texttt{contribution} (both on a 1--5 scale) as joint regressors:
\begin{equation}
P(\text{accept}) = \sigma(\alpha + \beta_S\cdot\text{soundness} + \beta_C\cdot\text{contribution}).
\end{equation}
For each dimension, its 50\% fitted threshold is computed while holding the other subscore at its within-group mean. High- and low-threshold topic groups are defined by the median $\hat{\tau}$ from the univariate rating model; two-sample tests compare estimated subscore thresholds between groups. Both recorded thresholds rise in the high-threshold group (Soundness 3.48 vs.\ 3.31; Contribution 3.52 vs.\ 3.29; both $p<0.05$). This diagnostic tests only a simple tradeoff on the recorded scales, which need not encode attack-specific criteria such as transferability and targeting~\citep{xu2025one}. It does not assume or establish that either subscore is comparable across topics, and it cannot exclude rational decisions based on area-specific meanings or unrecorded information.

\section{Confound and Robustness Tests}
\label{app:robustness}

\begin{figure}[t]
\centering
\includegraphics[width=\linewidth]{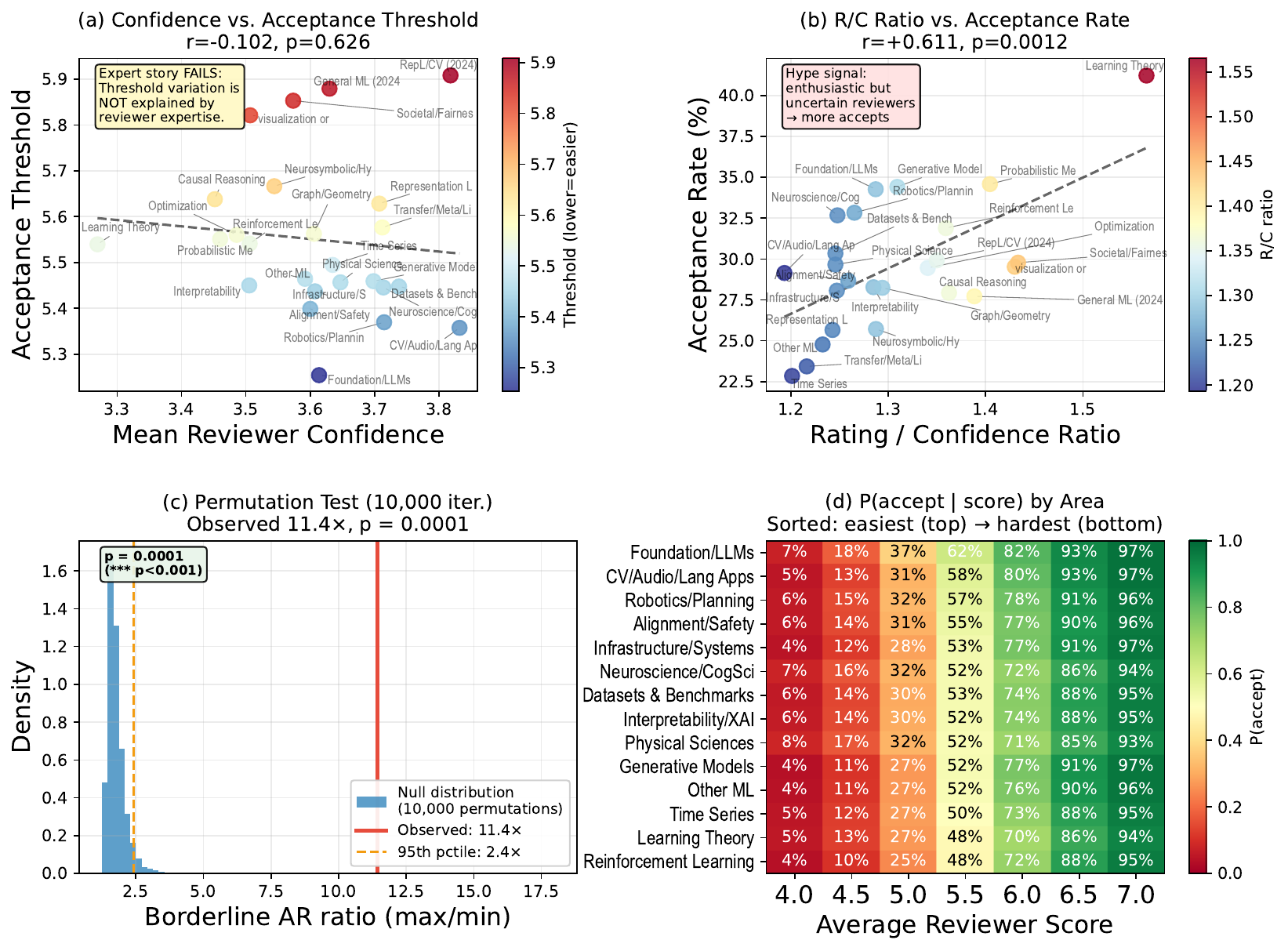}
\caption{\textbf{Robustness and confound analysis.}
Diagnostics for simple observable predictions of alternative explanations. (a)~Reviewer confidence vs.\ acceptance threshold: no significant correlation ($r=-0.102$, $p=0.626$). (b)~After within-area z-scoring, 79\% of the descriptive AR spread remains. (c)~Submission growth vs.\ AR change: no significant correlation ($r=-0.052$, $p=0.86$). (d)~Adding area to year and score-decile factors significantly improves fit. Panels (a)--(c) do not exclude more general latent-quality, expertise, or submission-composition models.}
\label{fig:robustness}
\end{figure}

\textbf{Panel (a): Expert-reviewer hypothesis.}
For each of 278 qualifying topics, we compute: (i) the per-topic acceptance threshold $\hat{\tau}$ from Appendix~\ref{app:logit}; and (ii) mean reviewer confidence (1--5). A simple expert-rigor explanation, together with the assumption that confidence measures expertise, predicts $r>0$. We obtain $r=-0.102$ ($p=0.626$, Pearson, two-tailed). A LOWESS smoother shows no clear nonlinear pattern. Because confidence is an imperfect proxy, this null result does not rule out expertise differences.

\textbf{Panel (b): Scoring-culture correction.}
To test a location--scale version of scoring culture, we z-score ratings within each area before computing acceptance rates. If differing means and variances entirely produced the gap, this transformation would remove it. The transformation leaves 79\% of the descriptive inter-area AR spread. This shows that a simple z-score correction is insufficient; it does not exclude nonlinear or area-specific scale meanings.

\textbf{Panel (c): Quality dilution hypothesis.}
For each of the 21 primary areas (2024--2026 taxonomy), we compute submission growth and change in AR over the same period. A simple mechanical-dilution model predicts a negative correlation. We obtain $r=-0.052$ ($p=0.86$); Datasets \& Benchmarks has $+459\%$ growth and a $-3.6$pp AR change. The test does not observe entrant quality and therefore cannot rule out compositional changes, including better entrants.

\textbf{Panel (d): ANOVA variance decomposition.}
A one-way ANOVA is run on binary acceptance outcomes with three nested factor sets: (i) year only; (ii) year + reviewer score decile; (iii) year + reviewer score decile + primary area. The incremental $F$-test for the addition of primary area is significant ($p < 0.01$), confirming that area identity explains acceptance variance beyond what score and year account for. The decomposition is offered as a complement to the logistic LRT in Appendix~\ref{app:logit}; both approaches converge on the same conclusion.

\section{NeurIPS 2024 External Validation Attempt}
\label{app:neurips}

\begin{figure}[t]
\centering
\includegraphics[width=1.0\linewidth]{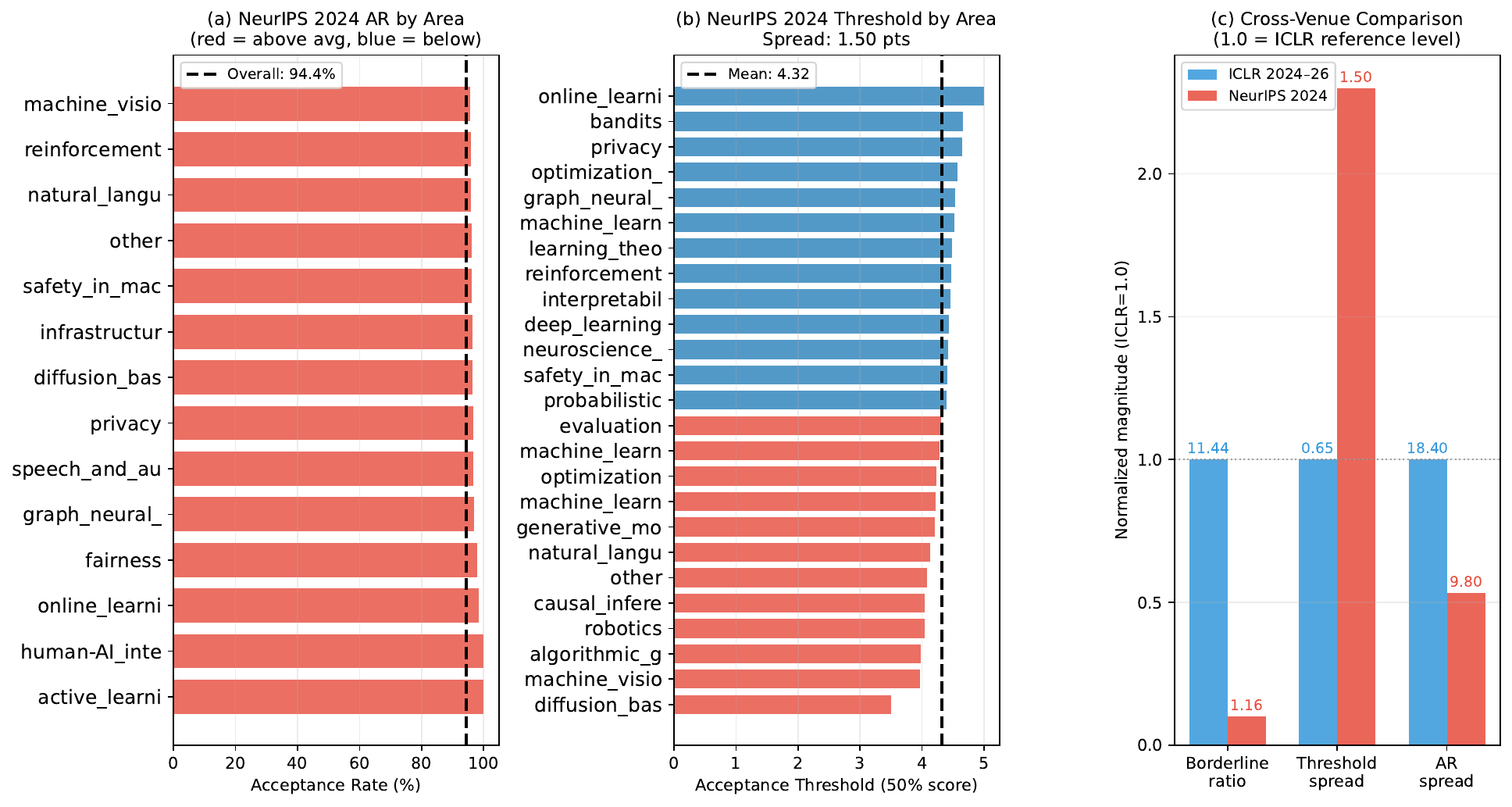}
\caption{\textbf{NeurIPS 2024 external-validation attempt (inconclusive).}
Public data are available for approximately 4,830 of $\sim$15,650 submissions. Because review disclosure is opt-in, the sample is dominated by accepted papers and contains only 268 rejected papers. This selection bias precludes a reliable score-conditional or external-validity conclusion. The figure documents the failed audit and provides no evidence for or against an ICLR-like effect.}
\label{fig:neurips}
\end{figure}

\textbf{Technical note.}
NeurIPS operates an opt-in review disclosure policy: authors must affirmatively choose to make their reviews public on OpenReview. This creates severe selection bias. Of approximately 15,650 submissions to NeurIPS 2024, roughly 4,830 reviews are publicly available. Critically, this available subset is dominated by accepted papers: we identify approximately 4,562 accepted papers with public reviews but only 268 rejected papers, a ratio of 17:1 in favor of accepts, compared to the true NeurIPS 2024 accept/reject ratio of approximately 1:4. This inversion means that any threshold or score-conditional AR analysis on the available NeurIPS data will dramatically understate the true acceptance bar and overstate acceptance rates for borderline papers.

The figure documents the attempt and the data limitation: the population-level audit used for ICLR is infeasible on this selected NeurIPS sample. We do not interpret visible threshold patterns as suggestive or confirmatory. Whether the effect is present at NeurIPS is unknown. The conference could answer the question by running the same audit on complete internal data or by providing a representative disclosure mechanism.

\section{Topic Taxonomy: Keyword Derivation and Paper Retrieval}
\label{app:topic_stats}

\subsection*{How Topic Keywords Are Derived}

Topics are derived from BERTopic~\citep{grootendorst2022bertopic} on title-and-abstract text for all 50,289 submissions. Each paper is encoded as a 3072-dimensional embedding using \texttt{text-embedding-3-large}; BERTopic applies UMAP (10 dimensions), HDBSCAN (\texttt{min\_cluster\_size=10}), and class-based TF--IDF labels. The initial run yields 322 non-outlier candidate labels. Cleaning and merging produces an exclusive 219-topic taxonomy used for the dataset overview and hard-assignment robustness check. For the main fine-grained analysis, deterministic regex retrieval retains the 322 candidate labels and permits overlapping memberships; 146 pass the forest-plot sample filters. Table~\ref{tab:topic_counts} distinguishes these counts.

\begin{table}[t]
\setlength{\aboverulesep}{0.2ex}\setlength{\belowrulesep}{0.2ex}
\centering
\begin{small}
\caption{Topic counts at each pipeline stage.}
\vspace{-0.1cm}
\label{tab:topic_counts}
\smallskip
\begin{tabular}{@{}>{\centering\arraybackslash}p{0.8cm}p{3.2cm}p{8.4cm}@{}}
\toprule
Count & Stage & Definition \\
\midrule
322 & Candidate overlapping labels & Initial non-outlier BERTopic labels retained for regex retrieval; a paper may match several. \\
219 & Cleaned exclusive taxonomy & Merged hard-assignment topics used for the dataset overview and an exclusive-topic robustness check. \\
278 & Threshold-eligible & Overlapping topics passing logistic-fit filters ($\hat{\beta}_1>0.3$, $\hat{\tau}\in[3.0,8.5]$); used in Figure~\ref{fig:threshold}. \\
146 & Main qualifying set & Overlapping topics with $\geq250$ regex-matched papers and $\geq30$ borderline papers; used in Figure~\ref{fig:systematic} and the one-row joint check. \\
\bottomrule
\end{tabular}
\end{small}
\vspace{-0.3cm}
\end{table}

\subsection*{How Papers Are Retrieved per Topic (Regex Matching)}

\begin{wraptable}{r}{0.22\textwidth}
\setlength{\aboverulesep}{0.2ex}\setlength{\belowrulesep}{0.2ex}
\vspace{-1ex}
\begin{center}
\begin{small}
\caption{Per-topic paper counts ($n=322$ topics, 2021--2026 pooled).}
\vspace{-2ex}
\label{tab:topic_sizes}
\smallskip
\begin{tabular}{cc}
\toprule
Statistic & Count \\
\midrule
Maximum & 3{,}637 \\
75th pctile & 421 \\
Median & 219 \\
Mean & 359.9 \\
25th pctile & 112 \\
Minimum & 31 \\
\bottomrule
\end{tabular}
\end{small}
\end{center}
\end{wraptable}

Each candidate topic's keyword list is converted into a deterministic regular-expression pattern applied to the lowercased concatenation of title and abstract (\texttt{re.IGNORECASE}). Patterns use tokens from the BERTopic-generated labels and are marked \texttt{source=``auto''} or \texttt{source=``override''} when manually corrected for over- or under-matching. One paper can match zero, one, or multiple topics. The pre-computed index contains 119,733 paper--topic pairs across 322 candidate topics and six years.

The table below summarizes the distribution of per-topic matched-paper counts pooled over 2021--2026. The spread from 31 to 3{,}637 reflects genuine variation in topic scope: broad topics such as ``graph neural networks'' match thousands of papers, while narrow methodological topics such as ``activation functions'' match a few hundred.

\subsection*{Comparison with Official \texttt{primary\_area} Labels}

The 21 official \texttt{primary\_area} labels (available only from 2024) provide a coarser author-assigned taxonomy. The BERTopic regex taxonomy spans all six years, uses paper content, and provides finer granularity through 322 candidate overlapping labels. These properties support longitudinal comparisons and detection of within-area score--acceptance variation.

\section{Rhetoric Analysis: Null Result}
\label{app:rhetoric}

\begin{figure}[t]
\centering
\includegraphics[width=1.0\linewidth]{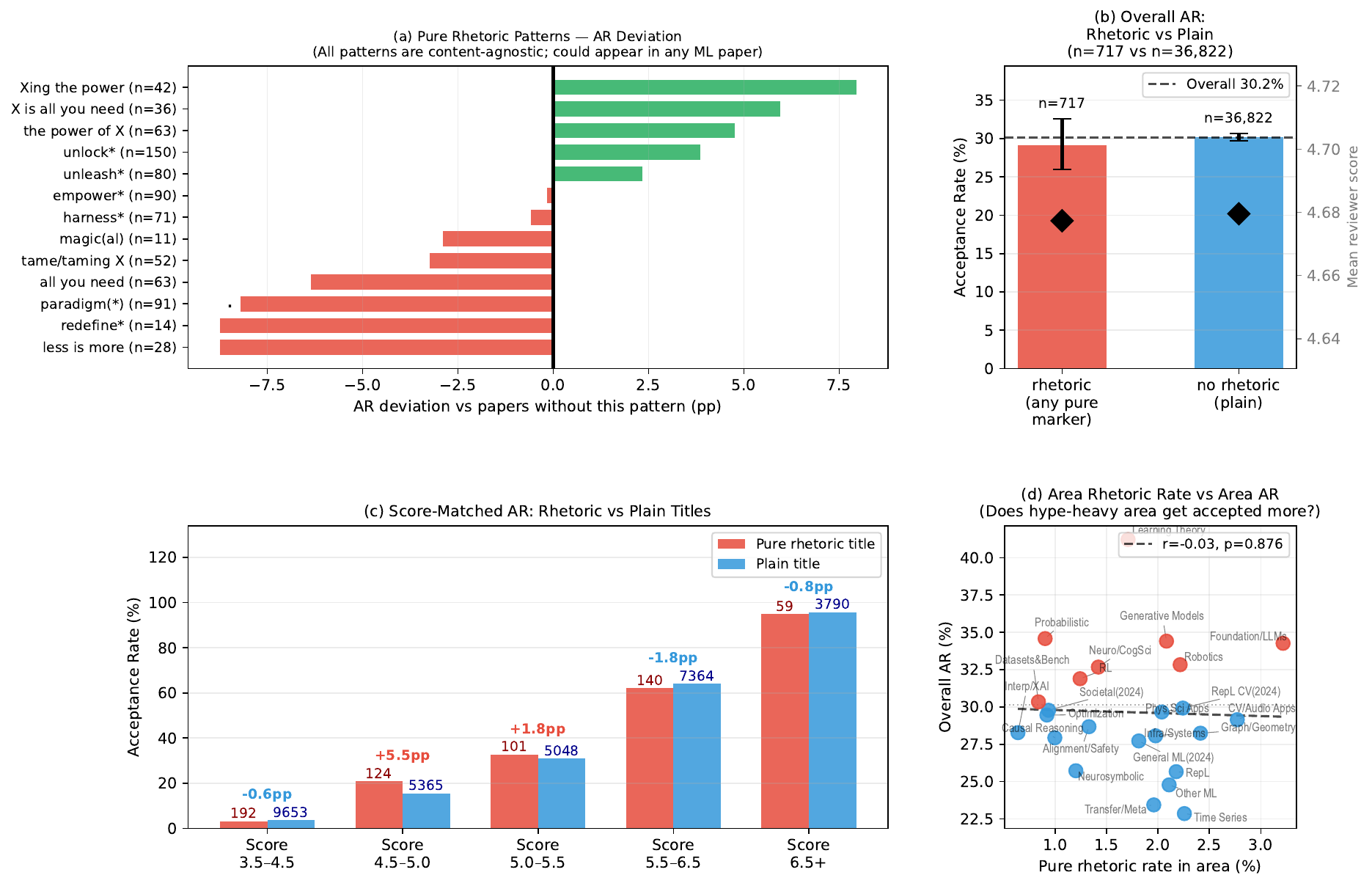}
\caption{\textbf{Promotional language analysis: null result.}
Using the \citet{millar2022hype} promotional language lexicon (139 words across 8 categories), we computed a promotional-language score for each paper title. Its correlation with acceptance rate across topics is near zero ($r\approx0.01$, $p=0.94$). This null diagnostic does not support a simple title-rhetoric explanation, but it cannot exclude other rhetorical effects.}
\label{fig:rhetoric}
\end{figure}

\textbf{Technical note.}
This analysis tests whether the observed AR disparities could be explained by rhetorical framing: papers in trending areas may use more promotional language in their titles, and reviewers may respond to promotional framing independently of paper quality. We use the promotional language lexicon from \citet{millar2022hype}, which comprises 139 words categorized into eight rhetorical strategies: superlatives (e.g., ``novel'', ``superior''), novelty claims (e.g., ``pioneering'', ``breakthrough''), hedges (e.g., ``promising'', ``potential''), quantitative claims, comparative claims, impact claims, methodological superlatives, and temporal claims. For each paper, we compute a title-level promotional language score as the fraction of title tokens matching any lexicon entry (after lowercasing and stop-word removal).

Per-topic mean promotional-language scores are correlated with per-topic acceptance rates (Pearson $r$, two-tailed $p$-value). The estimate is near zero ($r\approx0.01$, $p=0.94$). Abstract-level scores and score-residualized AR yield similarly null results. These diagnostics do not support the tested lexicon-based rhetoric explanation; they do not rule out broader differences in framing or writing.

\end{document}